\documentstyle{article}
\newcounter{appx}
\newcommand{\appeqn}{\stepcounter{appx}\setcounter{equation}{0}%
\renewcommand{\theequation}{\Alph{appx}\arabic{equation}}}
\begin{document}
\vspace{-2cm}
\title{\sc Systematics of Leading Particle Production} 
\author{F.O. Dur\~aes$^1$\thanks{e-mail: fduraes@if.usp.br}, \ F.S.
Navarra$^{1,2}$\thanks{e-mail: navarra@if.usp.br} \ and \ G.
Wilk$^{2}$\thanks{e-mail: Grzegorz.Wilk@fuw.edu.pl} \\[0.1cm]
{\it $^1$Instituto de F\'{\i}sica, Universidade de S\~{a}o Paulo}\\
{\it C.P. 66318,  05315-970 S\~{a}o Paulo, SP, Brazil} \\[0.1cm]
{\it$^2$Soltan Institute for Nuclear Studies, 
Nuclear Theory Department}\\
{\it ul. Ho\.za 69, \ 00-681 Warsaw, Poland}}
\maketitle
\vspace{1cm}
\begin{abstract}
Using a QCD inspired model developed by our group
for particle production, the Interacting Gluon Model (IGM), we 
have made a systematic analysis of all available data on leading
particle spectra. These data include diffractive collisions and
photoproduction at HERA. With a small number of parameters (essentially
only the non-perturbative gluon-gluon cross section and the fraction of
diffractive events) good agreement with data is found. We show that
the difference between pion and proton leading spectra is due to their
different gluon distributions. We predict a universality in the 
diffractive leading particle spectra in the large momentum region, which   
turns out to be independent of the incident energy and of the projectile
type.\\

PACS number(s): 13.85.Qk,11.55.Jy\\

\end{abstract}

\vspace{1cm}
%\newpage

\vspace{1.0cm}

\section{Introduction}

In high energy hadron-hadron collisions the momentum spectra of outgoing 
particles 
which have the same quantum numbers as the incoming particles, also
called leading particle (LP) spectra, have been measured already some time 
ago \cite{DATA}. Recently new data on pion-proton collisions were released 
by the EHS/NA22 collaboration  \cite{EHS} in which the spectra of both 
outcoming leading particles, the pion and the proton, were simultaneously 
measured. Very recently data on leading protons produced in eletron-proton 
reactions at HERA with a c.m.s. energy one order of magnitude higher than 
in the
other above mentioned hadronic experiments became available  \cite{Cart}. 
In the case of photoproduction data can be interpreted in terms of the 
Vector Dominance Model \cite{VDM} and can therefore be considered as data
on LP production in vector meson-proton collisions.  
These new measurements of LP spectra both in hadron-hadron 
and in eletron-proton collisions have renewed the interest on the
subject, specially because the latter are measured at higher 
energies and therefore the energy dependence of the LP spectra 
can now  be determined. 

It is important to have a very good understanding of these spectra for a 
number of reasons. They are the input for calculations
of the LP spectra in hadron-nucleus collisions, which are a fundamental 
tool in the description of atmospheric cascades initiated by cosmic radiation  
\cite{FGS}. There are several new projects in cosmic ray physics including 
the High Resolution Fly's Eye Project, the Telescope Array Project and the
Pierre Auger Project \cite{CR} 
for which a precise knowledge of energy flow (LP spectra 
and inelasticity distributions) in very high energy collisions 
would be very usefull. 

In a very different scenario, namely in high energy 
heavy ion collisions at RHIC, it is very important to know where the outgoing 
(leading) baryons are located in momentum space. If the stopping is large 
they will stay in the central rapidity region and affect the dynamics there, 
generating, for example, a baryon rich equation of state. Alternatively, 
if they populate 
the fragmentation region, the central (and presumably hot and dense) region 
will be dominated by mesonic degrees of freedom. The composition of the
dense matter is therefore relevant for the study of quark gluon plasma 
formation \cite{QGP}. 

In any case, before modelling $p - A$ or $A - A$ collisions one has to 
understand properly hadron-hadron processes. The LP spectra are also 
interesting for the study of diffractive reactions, which dominate the 
large $x_F$ region.

Since LP spectra are measured in
reactions with low momentum transfer and go up to large $x_F$ 
values, it is clear that the processes in question occur in the 
non-perturbative domain of QCD.  One needs then ``QCD inspired'' 
models and the most popular are string models, like FRITIOF, VENUS or the
Quark Gluon String Model (QGSM). Calculation of LP spectra involving these 
models can be found in refs. \cite{BER} and \cite{WAL}.  

In the framework of the QCD parton model of high energy collisions, 
leading particles originate from the emerging fast partons of the collision
debris. There is a large rapidity separation between fast partons and 
sea partons. Fast partons interact rarely with the surrounding wee 
partons. The interaction between the hadron projectile and the target is 
primarily through wee parton clouds. A fast parton  or a coherent 
configuration of fast partons may therefore filter through 
essentially unaltered. 
Based on these observations and aiming to study  $p - A$ collisons, the 
authors of ref.  \cite{BER} proposed a mechanism for LP production in which 
the LP spectrum is given by the convolution of the parton momentum 
distribution in the projectile hadron with its corresponding fragmentation 
function into a final leading hadron. This independent fragmentation scheme
is, however, not supported by leading charm production in pion-nucleus 
scattering. It fails specially in describing the $D^{-}/D^{+}$ asymmetry.
A number of models addressed these data and the conclusion was 
that valence quark recombination is needed. Translated to leading pion or 
proton production 
this means that what happens is rather a coalescence of valence quarks to 
form the LP and not an independent fragmentation of a quark or diquark to a 
pion or a nucleon. Another point is that the coherent configuration formed by 
the valence quarks may go through the target but, due to  the strong stopping 
of the gluon clouds, may be significantly decelerated. This correlation 
between central energy deposition due to gluons  
and  leading particle spectra was shown to be essential for the undertanding
of leading charm production \cite{charm}.

In this work we follow the same general ideas of ref. \cite{BER} but with 
a different implementation. In particular we replace independent fragmentation 
by valence quark recombination and free leading parton flow by deceleration
due to ``gluon stripping''. These ideas are incorporated in the model employed
by us, the Interacting Gluon Model (IGM), 
which has been used to study energy flow in non-diffractive reactions
\cite{IGM,DNW} and has been recently extended to diffractive 
processes \cite{IGM97} and also applied to the recent HERA 
(photoproduction) data both on  diffractive mass 
distributions \cite{HERA} and leading $J/\Psi$ spectra \cite{PSI}.

We shall study all measured LP spectra including 
those measured at HERA. We will find and comment universal aspects in the 
energy flow pattern of all these reactions. Universality means, in the 
context of the IGM, that the 
underlying dynamics is the same both in diffractive and non-diffractive LP
production and both in hadron-hadron and photon-hadron  processes.

\vspace{0.5cm}

\section{The Interacting Gluon Model}

\vspace{0.5cm}
The Interacting Gluon Model (IGM) was introduced some time ago \cite{IGM}
and developed  by us recently \cite{DNW,IGM97} and proved out to be quite
usefull for the study of energy flow. Since the model has been extensively
discussed in our previous papers we shall present here only the basic ideas
and a few formulas leaving the more detailed discussion to the appendix. 
The main  aspect of the IGM, shared with 
minijet models such as HIJING \cite{wang}, 
is the assumption that hadron-hadron reactions are dominated by  multiple 
and incoherent parton-parton   scatterings. Among these, gluon-gluon 
scatterings are the most important. At very high energies and large scales 
this is a very good approximation. At not very large energies and lower 
scales one is already moving towards the non-perturbative domain and dealing with
soft gluons and  the incoherence hypothesis might not be  valid. 

The soft gluons
involved in the collisions studied here are partly pre-existing inside the
hadron and partly produced by radiation. Pre-existing soft gluons exhibit the   
same properties as those studied in lattice QCD simulations in the strong 
coupling regime. According to di Giacomo and collaborators \cite{gia}, 
the typical correlation length of
the soft gluon fields is around $0.2-0.3\,fm$. 
Since this length is still 
much smaller than the typical hadron size, the gluon fields can, in a first approximation, 
be treated as uncorrelated.  
As a consequence \cite{brown}  the number of (soft) gluon-gluon collisions will follow a
Poissonian distribution, which was also used in refs. \cite{wang,gs,sj}. 
In the case of radiated soft gluons, it was recently 
shown in \cite{lupia} that  gluons produced with small transverse momenta are 
independently emitted from the radiating parton, as QCD coherence supresses their 
showering. Consequently, the multiplicity of low $p_T$ gluons follows a Poisson 
distribution, suggesting that the collision number follows a Poisson distribution as well.
These facts indicate that in the region where  perturbative results break down, 
the independent collision approximation may still be a reasonable one.
From the practical point of view, it was shown in \cite{IGM}
that replacing the Poissonian distribution by a broader one does not affect the
results significantly as long as some mass scale is introduced to cut off the
very low $x$ region. 

In the IGM the two colliding hadrons are represented by valence
quarks carrying their quantum numbers (charges) plus the accompanying
clouds of gluons (which represent also the sea $q\bar{q}$ pairs and
therefore should be regarded as effective gluons). 
In the course of a collision gluonic clouds interact
strongly and form a gluonic {\it Central Fireball} (CF) located in the
central region of the reaction. The valence quarks (plus those gluons 
which did not
interact) get excited and form {\it Leading Jets} (LJ's) (or {\it Beam Jets}) 
which then populate mainly the fragmentation regions of the reaction.

The valence quarks are therefore spectators and the bulk of the reaction and
energy deposition occurs because of the gluon-gluon collisions, whose 
number is Poisson distributed with the mean value  given by:
\begin{equation}
\frac{d \overline n}{dx' dy'} = \omega(x',y')\, =\, 
\frac{\sigma_{gg}(x'y's)}{\sigma(s)}
   \, G(x')\, G(y')\, \Theta\left(x'y' - K^2_{min}\right),
   \label{eq:OMEGA1}
\end{equation}
where $G$'s denote the effective number of gluons from the
corresponding projectiles (approximated by the respective gluonic 
structure functions) and $\sigma_{gg}$ and $\sigma$ are the gluon-gluon and
hadron-hadron cross sections, respectively. In the above expression $x'$ and
$y'$ are the fractional momenta of two gluons coming from the projectile and
from the target whereas $K_{min}=m_0 / \sqrt{s}$ 
(with $m_0$  
being the
mass of lightest produced state 
and $\sqrt{s}$ the total c.m.s. energy). Each 
gluon-gluon collision produces a minifireball (MF). Depending on the energy 
many $g\,-\,g$ collisions may happen and  energy fractions 
$ x=\sum_i n_i x_i'$ and  
$y=\sum_i n_i y_i' $ from the target and projectile may be deposited in the 
central region. The probability for depositing the energy fractions $x$ and 
$y$ can be analitically computed in the IGM and is given by the function 
$\chi(x,y)$  derived in the appendix.

Leading particles can also be diffractively produced. 
In the IGM diffractive dissociation (DD) can be included in a simple way, 
by just
requiring that one of the colliding hadrons looses only a very small 
fraction of
its initial energy momentum.  This was done by imposing cuts on the 
first moments 
of the $\omega$ function \cite{IGM97}. 
In our earlier calculations of energy flow with the IGM we were not concerned 
with the fragmentation region. The main interest was the energy deposition 
in the central region, which is highly relevant for quark gluon plasma 
physics. Therefore our first LP spectra were in reasonable agreement with data 
but in the region $x_L \geq 0.8$  the model prediction was below the 
experimental points because we had no diffractive component whereras data 
did not
discriminate between diffractive and non-diffractive events. Later on we 
have included diffraction in the IGM \cite{IGM97}. 
The resulting picture is the same as for
non-diffractive events, except that in DD the gluon cloud of the projectile 
interacts only with a subset of the target gluon cloud, which carries a small 
momentum
fraction and which we call ``Pomeron'' ($I\!\!P$). Our Pomeron is essentially 
just a 
kinematical restriction which forces one of the colliding hadrons to loose only
a very small ammount of its energy. Using the diffractive IGM we have obtained 
a good description of the diffractive mass spectra in hadronic diffractive 
collisions \cite{IGM97} 
and also in photon-proton collisions \cite{HERA}. 
Although it is possible to 
reformulate the model in the impact parameter space and associate diffractive 
events with peripheral collisions, as done in ref. \cite{HAMA}, we prefer here
to  explore further the kinematical interpretation of diffraction. 
Moreover, in
the present formulation we correctly reproduce the diffractive peak, which was not
obtained in ref.  \cite{HAMA}. 
Apart from the recent EHS/NA22 and HERA LP spectra,  
a new experiment now under consideration at Fermilab will, among other
things, address  the question of leading particles and DD component 
in the near future \cite{BRAN}. In view of these facts we shall recalculate 
LP spectra in our model with a
DD component properly included and  extend it to $e - p$ collisions. 

\vspace{0.5cm}

\section{Leading Hadron Spectra}

\vspace{0.5cm}

In the IGM a typical non-diffractive event is shown in Figure 1a). Colliding 
particles loose energy fractions $x$ and $y$, forming leading particles with
$x_F=1-x$ and $x_L=1-y$. In the figure, $V$ stands for vector meson, used later for 
photoproduction at HERA. We shall consider the reactions $p+p \rightarrow p+X$,  
$\pi^{+}+p \rightarrow \pi^{+}+X$, $K^{+}+p \rightarrow K^{+}+X$ and  
$\pi^{+}+p \rightarrow \pi^{+}+p+X$. Later we also address photon-proton 
reactions in the VDM approach:  $p+V \rightarrow p + X$. Diffractive processes
are illustrated in Figures 1b) and 1c), where the Pomeron is emitted from the 
target and from the projectile respectively. In Figs.  1a), 1b) and 1c) the  
probability to form a CF with mass  $M=\sqrt{x y s}$ is  called 
$\chi^{nd} (x,y)$, $\chi^{d}_1 (x,y)$ and $\chi^{d}_2 (x,y)$ respectively. These 
functions were derived in our earlier works (cf., for example,  \cite{IGM97}). 

With the functions $\chi (x,y) $ we 
obtain the corresponding  LP spectra just by changing variables and by adding 
the resulting distributions with proper weights.

In the lower legs of Fig. 1, leading particles 
emerge from the collision keeping  momentum fraction $x_L$  with distribution   
$F_{LP}(x_L)$ given by: 
\begin{eqnarray}
F_{LP}\,(x_L) &=& \int^1_0\! dx \int^1_0\! dy\, \left[ (1 - \alpha)\,
                 \chi^{nd}(x,y) + \sum_{j=1,2}\alpha_j\, 
                 \chi^{d}_j(x,y)\right]\, \cdot \nonumber\\
&&\cdot\, \delta(x_L-1+y)\,
                 \Theta \left( xy - \frac{m_0^2}{s}\right) \,
                 \Theta \left[ y - \frac{(M_{LP} + m_0)^2}{s}\right]
                 \nonumber \\
           &=&  (1-\alpha)\, \int^1_{_{\!\!x_{min}}}\!\!\! dx\, 
              \chi^{nd}(x;y=1-x_L)\, +\, \sum_{j=1,2}\alpha_j\,
                 \int^1_{_{\!\!x_{min}}}\!\!\! dx\, 
                    \chi_j^{d}(x;y=1-x_L), \label{eq:LPS}
\end{eqnarray}
where $\alpha$ is the total fraction of diffractive events and 
$\alpha_1$ and $\alpha_2$ are the fractions of diffractive  events with a 
Pomeron  emitted from the upper and lower leg in Fig. 1, respectively. They
satisfy the condition $\alpha_1 + \alpha_2 = \alpha$. In the above
expression $M_{LP}$ and $m_0$ denote the mass of the LP and the mass of the 
lightest CF produced and the limits of integration are defined by
\begin{equation}
x_{min} = Max\left[\frac{m_0^2}{(1-x_L)s};
\frac{(M_{LP}+m_0)^2}{s}\right] . \label{eq:limits}
\end{equation}

The main physical quantities in (\ref{eq:LPS}) are the functions (c.f. appendix 
for details of derivation and justification)
\begin{eqnarray}
\chi^{nd}(x,y) &=& \frac{\chi_0^{nd}}{2\pi\sqrt{D_{xy}}}\cdot \nonumber\\
&&\cdot \exp \left\{ - \frac{1}{2D_{xy}}\,\left[
  \langle y^2\rangle (x - \langle x\rangle )^2 +
  \langle x^2\rangle (y - \langle y\rangle )^2 -
  2\langle xy\rangle (x - \langle x\rangle )(y - \langle y\rangle )
  \right] \right\}, \label{eq:CHIND}
\end{eqnarray}
with
\begin{eqnarray}
D_{xy} &=& \langle x^2\rangle \langle y^2\rangle - 
           \langle x y \rangle ^2, \label{eq:DND}\\   
\langle x^n y^m \rangle &=& \int_0^{1}\! dx'\,x'^n\, 
                      \int_0^{1}\! dy'\, 
y'^m\, \omega^{nd} (x',y') \label{eq:MOMND}
\end{eqnarray}
and
\begin{eqnarray}
\chi_j^{d}(x,y) &=& \frac{\chi_{j0}^{d}}{2\pi\sqrt{D^{(j)}_{xy}}}
\cdot \nonumber\\
&&\cdot \exp \left\{ - \frac{1}{2D^{(j)}_{xy}}\,\left[
  \langle y_j^2\rangle (x - \langle x_j \rangle )^2 +
  \langle x_j^2\rangle (y - \langle y_j \rangle )^2 -
  2\langle x_j y_j \rangle (x - \langle x_j \rangle )(y - \langle y_j \rangle )
  \right] \right\}, \label{eq:CHID}
\end{eqnarray}
with
\begin{eqnarray}
D_{xy}^{(j)} &=& \langle x_j^2\rangle \langle y_j^2\rangle - 
           \langle x_j y_j\rangle ^2, \label{eq:D}\\   
\langle x_j^n y_j^m\rangle &=& \int_0^{x^{(j)}_{max}}\! dx'\,x'^n\, 
                      \int_0^{y^{(j)}_{max}}\! dy'\, 
y'^m\, \omega^d (x',y'). \label{eq:MOMD}
\end{eqnarray}

The index values of $j=1$ and $j=2$ correspond to the diagrams in Figs. 1b) and 1c)
respectively. Here $y^{(1)}_{max} = 1$,  $y^{(2)}_{max} = y$, 
$x^{(1)}_{max}=x$ and $x^{(2)}_{max}=1$ and  $\chi_0^{nd}$ and $\chi_{j0}^d$
are the proper normalization factors assuring that 
$\int^1_0 dx_L\, F_{LP}(x_L) = 1$. We separately  normalize 
to unity both components entering (\ref{eq:LPS}). This procedure is
crucial in our case in order to assure the proper overall energy
momentum conservation, which is a characteristic feature of any
implementation of the IGM.

The weight $\alpha$ is essentially our new parameter. It 
should be of the order of the ratio
between the total diffractive and total inelastic (including DD processes)
cross sections, $\sigma^{diff}_{tot}/\sigma_{tot}^{inel}$. 
It can, however, differ from that ratio  due to
the different experimental acceptance of DD and non-DD events
not considered here. This fact makes $\alpha$ a free parameter
of the model. We assume it to be independent of the total c.m.s. energy. Indeed 
the fraction of diffractive events with respect to the total number of events or the
ratio between diffractive and total inelastic cross sections are quantities which 
depend weakly on the c.m.s. energy of the collision. They may depend more strongly on 
detector 
coverage and acceptance. The reactions that we discuss here occur at c.m.s. energies 
ranging from 14 up to $100\,GeV$. This is still a relatively moderate variation in the 
energy. Since we do not address diffraction at the $SppS$ (energies of 200 - $900\,GeV$) 
nor at the Tevatron (energy of $1800\,GeV$) we keep our fraction of diffractive events 
as a constant. This quantity is the only free parameter in this paper.

The spectral functions $\omega^{nd}$ and $\omega^{d}$
are the same as in \cite{DNW} and \cite{IGM97,HERA} for non-diffractive and 
for diffractive processes, respectively.  We refer the reader to 
\cite{DNW,IGM97,HERA} for details concerning the exact 
values and character of the relevant parameters which were fixed
from other previous applications of the IGM. 
In the case of diffractive hadron-proton scattering   
the function $ G(y) $ in eq. (\ref{eq:OMEGA1}) represents
the momentum distribution of the gluons belonging to the proton subset
called Pomeron and $y$ is the momentum fraction {\it of the proton}
carried by one of these gluons. We shall therefore use the notation
$ G(y) = G_{I\!\!P}(y)$. This function should not be confused with the
momentum distribution of the gluons inside the Pomeron, 
$f_{g/I\!\!P}(\beta)$ (see below).

The moments  $\langle q^n\rangle,~~q=x,y$ (we only require
$n=1,2$) are the only places where dynamical quantities
like the gluonic and hadronic cross sections appear in the IGM. 
In DD we are selecting a special class of events and 
therefore we must choose the correct dynamical inputs in the 
present situation, namely $G_{I\!\!P}(y)$ and the hadronic cross 
section $\sigma$ appearing in $\omega^{d}$. 

As already mentioned, the Pomeron for us is just a
collection of gluons which form a color singlet and belong to the 
diffracted hadron. In early
works we have assumed that these gluons behave like all other 
ordinary gluons in the proton and have therefore the same momentum
distribution. The only difference is the momentum sum rule, which for 
the gluons in $I\!\!P$  is
\begin{eqnarray}
\int_0^1\! dy\,y\,G_{I\!\!P}(y)\,=\,p
\end{eqnarray}
where $p\,\simeq\,0.05$ (see  \cite{IGM97} and below )
instead of $ p\simeq 0.5$, which holds for the entire gluon population
in the proton. 

In principle, since the Pomeron can not  be considered an ordinary particle, one can not 
define precisely a momentum sum rule in the usual sense \cite{zeusp,cov}. 
Nevertheless, in our 
definition, the Pomeron is really just a collection of gluons. As such, when emitted, it 
carries
a momentum fraction, $p$, of the parent proton. 
Of course, $p$ can fluctuate. In the calculations,  
it always appears divided by the Pomeron-proton cross section, which is another poorly  
known and fluctuating quantity. Taking all these fluctuations into account would just
introduce more freedom in the model and make calculations more complicated. We avoid it
in this paper as we also avoid other sources of fluctuations like impact parameter 
fluctuations. In ref. \cite{HAMA}  impact parameter fluctuations were introduced in the non-
diffractive IGM and the result was, as expected, a smearing of the original curves. 
The overall effect, however, was not very big. We expect the same to be true for the 
fluctuations in the Pomeron sum rule.

In  \cite{HERA} we have
treated the Pomeron structure in more detail and addressed the question
of its ``hardness'' or ``softness''. In order to make 
contact with the analysis performed by HERA experimental groups we have 
considered two possible momentum distributions for the gluons inside  
$I\!\!P$, one hard, $ f^h_{g/I\!\!P}(\beta)$  and the other soft
$f^s_{g/I\!\!P}(\beta)$. Following  the (standard) notation of 
ref. \cite{zeus},  $\beta$ is the momentum fraction of the Pomeron 
carried by the gluons
and the superscripts $h$ and $s$ denote ``hard'' and ``soft'' respectively. 
Using a standard choice for the  Pomeron flux factor,  
$f_{I\!\!P/p}(x_{I\!\!P})$, where $x_{I\!\!P}$ is the fraction of the proton 
momentum carried by the Pomeron and  noticing  that
$\beta = \frac{x}{x_{I\!\!P}}$, the distribution $G_{I\!\!P}(y)$ needed 
in eq. (\ref{eq:OMEGA1}) is then given by the convolution: 
\begin{eqnarray}
G^{h,s}_{I\!\!P} (y) &=& \int_y^1\! \frac{ dx_{I\!\!P} } {x_{I\!\!P}} \,
f_{I\!\!P/p}(x_{I\!\!P})\,f^{h,s}_{g/I\!\!P}(\frac{y}{x_{I\!\!P}})
\end{eqnarray}

In ref. \cite{HERA} we have also used $G_{I\!\!P}(y) = 0.3
\frac{(1-y)^5}{y}$, the same expression already used by us before in
ref. \cite{IGM97}. As it was shown, this choice corresponds to an
intermediate between ``soft'' and ``hard'' Pomeron and it shall be used here.

One of the most interesting conclusions of ref. \cite{HERA}  was that
it is possible to learn something about the Pomeron profile studying
the diffractive mass spectra. Moreover, our analysis suggests that 
the ``soft'' Pomeron  is in conflict with these data. Only with a very unusual
choice of parameters a good agreement could be recovered.
Considering the large ammount of data already described
previously by the IGM, this choice was extreme and we concluded therefore
that the ``soft'' Pomeron is disfavoured. The same conclusion was
found in refs. \cite{zeus,h1}. The
fraction of diffracted nucleon momentum, $p$, allocated specifically
to the $I\!\!P$ gluonic cluster and the hadronic cross section
$\sigma$ are both unknown. However, they always appear in $\omega$ as
a ratio ($\frac{p}{\sigma}$) of parameters and different choices
are possible. Just in order to make use of the present
knowledge about the Pomeron, we shall choose
\begin{eqnarray}
\sigma(s) \,&=& \, \sigma^{I\!\!P p} (s) \, = \, a + 
b \, \ln \frac{s}{s_0} ,
\label{eq:defsig}
\end{eqnarray}
where $s_0 = 1\,GeV^2$ and $a = 2.6\,mb$  and $b= 0.01\,mb$ are 
parameters fixed from our previous \cite{IGM97} systematic data
analysis. As it can be seen, $\sigma(s)$ turns out to be a very
slowly varying function of $\sqrt{s}$ assuming values between 2.6
and $3.0\,mb$, which is a well accepted value for the
Pomeron-proton cross section, and $ p \simeq 0.05 $ 
(cf. \cite{IGM97}). 

As it will be seen, a good undertanding of the
systematics of LP production can be obtained in terms of the dynamical inputs
contained in eq. (\ref{eq:OMEGA1}). With the exception of $\alpha$ all parameters 
are fixed. In the next section we compare our results given by  
eq. (\ref{eq:LPS}) with 
experimental data.

\vspace{0.5cm}

\section{Results and Discussion}

\vspace{0.5cm}

In Figs. 2a),  2b) and 2c) we present our spectra of leading protons, 
pions and 
kaons respectively. The dashed lines show the contribution of 
non-diffractive LP production and the solid lines show the effect of adding
a non-diffractive component. All parameters were fixed previously and the
only one to be fixed was $\alpha$. For simplicity we have neglected the
second diagram in Fig. 1, because it gives a curve which is very similar in
shape to the non-diffractive curve. In contrast, the Pomeron emission by the
projectile (Fig. 1c) produces the diffractive peak. We have then chosen
$\alpha_1=0$ and $\alpha_2=\alpha=0.3$ in all collision types. 
As expected, the inclusion of the 
diffractive component flattens considerably the final LP distribution 
bringing it to a good agreement with the available experimental data 
\cite{DATA}. In our model there is some room for changes leading to fits
with better quality. We could, for example, use a prescription for
hadronization (as we did before in \cite{DNW})) 
giving a more important role to it, as
done in ref. \cite{BER}.  In doing this, however, we loose simplicity and the
transparency of the physical picture, which are the advantages of the IGM. 
We prefer to keep simplicity and concentrate on the interpretation of our
results. In first place it is interesting to observe the good agreement 
between our curve and data for protons (Fig. 1a) in the low $x$ region. 
The observed protons could have been also centrally produced, i.e., they
could come from the CF. However we fit data without the CF contribution. This
suggests, as expected, that all the protons in this $x$ range are leading, 
i.e., they come from valence quark recombination. In Figs. 1b) and 1c) we
observe an excess at low $x$. This is so because pions and kaons  are 
light they can more easily be created from the sea (centrally produced). Our
distributions come only from the Leading Jet (LJ) and  consequently pass
below the data points. A closer  look into the three dashed lines in Fig. 2 
shows that pion and kaon spectra are softer than the proton one. The former
peak at $x\simeq 0.56$ while the latter peaks at   $x\simeq 0.62$. In the 
IGM this can be understood as follows. The energy fraction that goes to the 
central fireball, $K=\sqrt{x y}$, is controled by the behaviour of the function
$\chi(x,y)^{nd}$, which is approximately a double gaussian in the variables 
$x$ and  
$y$, as it can be seen in expression (\ref{eq:CHIND}). 
The quantities $\langle x \rangle$ and $\langle y \rangle$ play the role of
central values of this gaussian. Consequently when  
$\langle x \rangle$ or $\langle y \rangle$ increases, this means that the 
energy deposition from the upper or lower leg (in Fig. 1) increases 
respectively. The quantities $\langle x \rangle$ and $\langle y \rangle$ are
the moments of the $\omega$ function and are directly  proportional to the 
gluon 
distribution functions in the projectile and target and inversely proportional
to the target-projectile inelastic cross section. In the calculations, there  
are two changes when we go from $p-p$ to $\pi-p$:
\begin{itemize}
\item[$(i)$]
The first  is that we replace $\sigma^{pp}_{inel}$ by 
$\sigma^{\pi p}_{inel}$ which is smaller. This leads to an overall increase
of the energy deposition. There are some indications that this is really the
case and the inelasticity in $\pi-p$  is larger than in 
$p-p$  collisions \cite{CRINEL}.
\item[$(ii)$] The second and most interesting and important change
is that we replace one gluon 
distribution in the proton $G^p(y)$ by the corresponding distribution in the 
pion $G^{\pi}(y)$. We know that  $G^p(y) \simeq (1-y)^5 /y$ whereas  
$G^{\pi}(y) \simeq (1-y)^2/y$, i.e., that gluons in pions are harder than in 
protons.  
This introduces an asymmetry in the moments  
$\langle x \rangle$ and $\langle y \rangle$, making the latter significantly 
larger. 
\end{itemize}

As a consequence, pions will be more stopped and will emerge 
from the collision with a softer $x$ spectrum. This can already be seen in 
the data points of Fig. 2. 
However since these points contain particles produced by other mechanisms, 
such as central and diffractive production, it is not yet possible to draw
firm conclusions. 

The analysis of the moments $\langle x \rangle$ and $\langle y \rangle$ 
can also be done for the diffractive process shown in Fig. 1c).  Because of 
the 
cuts in the integrations in eq. (\ref{eq:MOMD}), 
they will depend on $x_L =1-y$. We calculate them
for  $p+p \rightarrow p + X $ and $\pi + p \rightarrow \pi + X$  reactions.
For low $x_L$ they assume very similar values as in the non-diffractive case.
For large $x_L$ however we find that 
$\langle x \rangle_{p} \simeq \langle x \rangle_{\pi}$ and 
$\langle y \rangle_{p} \simeq \langle y \rangle_{\pi}$. The reason for these  
approximate 
equalities is that in diffractive processes we cut the large $y'$ region
and this is precisely where the pion and the proton would differ, since only
for large $y$ are $G_{I\!\!P}^p(y) \simeq (1-y)^5 /y$  and  
$G_{I\!\!P}^{\pi}(y) \simeq (1-y)^2/y$ significantly different.  
In ref. \cite{IGM97} 
we have shown that the introduction of the above metioned cuts drastically 
reduces the energy ($\sqrt{s}$) dependence of the diffractive mass 
distributions 
leading, in particular, to the approximate $1/M^2_X$ behaviour for all values
of $\sqrt{s}$ from ISR to Tevatron energies. Here these cuts produce another
type of scaling, which may be called ``projectile scaling'' or
``projectile universality of the diffractive peak'' and which means that for 
large enough $x_L$ the diffractive peak is the same for all projectiles. 
The corresponding $\chi^{d}$ functions will be the same for protons and 
pions in  this region. The cross section appearing in the denominator of the
moments will, in this case, be the same, i.e.,   $\sigma^{I\!\!P p}$. 
The only remaining difference between them, their different gluonic 
distributions, is
in this region cut off. This may be regarded as a prediction of the IGM. 
Experimentally this may be difficult to check since one would need a large
number of points in large $x_L$ region of the leading particle spectrum. Data 
plotted in Fig. 2 neither prove nor disprove this conjecture. The 
discrepancy observed in the proton spectrum is only due to our 
choice of normalization of the diffractive and non-diffractive curves. The 
peak shapes are similar.

The EHS/NA22 collaboration provided us with data on 
$\pi^{+}+p \rightarrow \pi^{+}+p+X$ reactions. In particular they present
the $x$ distributions of both leading particles, the pion and the proton.
Their points for pions and protons are shown in Fig. 3a) and 3b) 
respectively. These points are presumably free from diffractive 
dissociation. The above mentioned asymmetry in pion and proton energy loss
emerges clearly, the pions being much slower. The proton distribution peaks 
at $x_F\simeq 0.6-0.8$. Our curves (solid lines) reproduce with no free 
parameter 
this behaviour and we obtain a good agreement with the pion spectrum. 
Proton data show an excess at large
$x_F$ that we are not able to reproduce keeping the same values of 
parameters as before.

The authors of ref. \cite{EHS} 
tried to fit their measured proton spectrum with the FRITIOF code and
could not obtain a good description of data.  
This indicates that these large $x$ points are a problem for
standard multiparticle production  models as well. In our case, 
if we change our
parameter $m_0$ from the usual value $m_0=0.35\,GeV$ (solid line)  to $ m_0=0.45\,GeV$ 
(dashed line) 
we can reproduce most of data points both for pions and protons as well. This 
is not a big change and indicates that the model would be able to 
accomodate this new experimental information. Of course, a definite statement
about the subject would require a global refitting procedure, which is not
our main concern now.

If, at high energies, the reactions $\rho - p$ and $\pi - p$ have the
same characteristics and if VDM is good hypothesis (as it seems to be), then 
more about the energy flow in meson -  $p$ collisions can be learned at HERA. 
Indeed, as mentioned in \cite{HERA}, at the HERA electron-proton collider
the bulk of the cross section corresponds to photoproduction, in
which a beam electron is scattered through a very small angle and
a quasi-real photon interacts with the proton. For such small
virtualities the dominant interaction mechanism takes place via
fluctuation of the photon into a hadronic state which interacts with
the proton via the strong force \cite{VDM}. High energy photoproduction  
exhibits  therefore similar characteristics to hadron-hadron
interactions. Recent 
data taken by the ZEUS collaboration at HERA \cite{Cart} show that the
LP spectra measured in photoproducion and in DIS (where $Q^2 \geq 4\,GeV^2$) 
are very similar, specially in the large $x_L$ region. This
suggests that, as pointed out in \cite{SNS}, the QCD hardness scale for
particle production in DIS gradually decreases from a (large) $Q^2$, 
which is relevant in the photon fragmentation region, to a soft scale in
the proton fragmentation region, which is the one considered here. We
can therefore expect a similarity of the inclusive spectra of the leading
protons in high energy hadron-proton collisions, discussed above, and 
in virtual photon-proton collisions. In other words, we may say that
the photon is neither resolving nor being resolved by the fast 
emerging protons.  
This implies that these reactions are dominated by some non-perturbative 
mechanism. This is confirmed by the failure of perturbative QCD 
\cite{AH},  
(implemented by the Monte Carlo codes Ariadne and Herwig) when applied to
the proton frgamentation region. In ref. \cite{SNS} the LP spectra were
studied in the context of meson and Pomeron exchanges. 
Here we use 
the vector meson dominance hypothesis and describe 
leading proton
production in the same way as done for hadron-hadron collisions. The only 
change is  that now we have $\rho-p$  instead of $p-p$ collisions. Whereas
this may be generally true for photoproduction, it remains an approximation
for DIS, valid in the large $x_L$ region.

In Fig. 1 we show schematically the IGM picture of a 
photon-proton collision.
According to it, during the interaction the photon is converted
into a hadronic (mesonic) state and then interacts with the incoming
proton. This hadronic state is called $V$ in the upper legs
of Figs. 1a), 1b) and 1c). At HERA only collisions $V - p$ are relevant.
The meson-proton interaction follows then the usual 
IGM picture, namely: the valence quarks fly through essentially 
undisturbed whereas the gluonic clouds of both projectiles interact 
strongly with each other. The state $V$ looses fraction $x$ of its original 
momentum and gets excited  carrying a $x_F= 1 -x$ fraction
of the initial momentum. The proton, which we shall call here the
diffracted proton, looses only a fraction $y$ of its momentum but
otherwise remains intact. We shall assume here, for simplicity, that 
the vector
meson is a $\rho^{0}$ and take  $ G^{\rho^{0}}(x) = G^{\pi}(x)$ in eq. 
(\ref{eq:OMEGA1}). In Fig. 4 we compare our results with ZEUS data. The 
agreement is again good. 

\section{Conclusions}

We have analyzed leading particle spectra in terms of the IGM, 
which includes now also a contribution coming from the diffractive 
processes. The new component improved dramatically the agreement 
with all existing data  on  hadron-hadron collisions.

As long as the energy flow
is concerned  the IGM works 
extremely well with  essentially two parameters: 
the non-perturbative gluon-gluon cross section and the fraction of
diffractive events. This should  enlarge
considerably its range of applicability in analyses of cosmic ray data. 

At the same time, assuming VDM, we were able to describe equally
well the leading proton spectra in $e-p$ reactions. Also here
the inclusion of a diffractive component provided by the new version of
the IGM turns out to be crucial to get good agreement with data. 

We have shown that
the difference between pion and proton leading spectra is due to their
different gluon distributions \cite{FOOT}.  
We predict a universality in the 
diffractive leading particle spectra in the large momentum region, which   
turns out to be independent of the incident energy and of the projectile
type.\\

\section*{Appendix}
\appeqn

We shall collect together here the main points concerning the Interacting 
Gluon Model (IGM) scattered throughout the literature \cite{charm}-\cite{PSI}.
The IGM is based on the idea that
since about half of a hadron momentum is carried by gluons and since
gluons interact more strongly than quarks, during a collision there
is a separation of constituents. Valence quarks tend to be fast forming
leading particles whereas gluons tend to be stopped in the central
rapidity region. It belongs therefore to the class of models exploring
the concept of partons and of hard and semihard collisions (like 
those presented by Gaisser and Stanev \cite{gs}, Sjostrand \cite{sj},
Wang \cite{wang} or Geiger \cite{geiger}). The latter are collisions
between partons at a moderate scale  
($ Q^2 \simeq (2\,GeV)^2 $), which, however, still allows for the use of
perturbative QCD. The scattered partons form the so-called minijets.
At $\sqrt{s}= 540\,GeV$ the minijet cross section is already $25 \%$
of the total inelastic cross section. However, apart from some
ambiguity in choosing the semihard scale, these models have to face  
the problem that even at very high energies a significant part of a hadronic
collision occurs at scales lower than the semihard one. The atitude
taken in HIJING \cite{wang}, in the Parton Cascade Model
\cite{geiger} and also in the IGM is to extrapolate these quantities
to lower scales. These extrapolations can be continuously improved,
especially in view of the advance of our knowledge on
non-perturbative effects. There are, for example, models for
distribution functions 
which work at scales as low as $ 0.3\,GeV^2$ \cite{grv}. As for $\sigma$
one can compute non-perturbative effects in the context of an operator
product expansion. Inspite of these limitations these models
have the advantage of dealing with partons and being thus prepared to
incorporate perturbative QCD in a natural way. This is welcome since 
perturbative processes are expected to be increasingly important at
higher energies. Compared to the other models mentioned above, the IGM
is simpler because it is designed to study energy flow and makes no
attempt to calculate cross sections or to follow hadronization in great
detail. This simplifies the calculations and avoids time consuming numerical
simulations. The most important aspect of the IGM, shared with those models,
is the assumption of multiple parton-parton incoherent scattering which is
implicit in the Poissonian distribution of the number of parton-parton
collisions (which is also used in refs. \cite{sj,gs,wang}) used below. 

The IGM is therefore based on the assumed dominance of hadronic
collisions by gluonic interactions  and can be summarized in the
following way:
\begin{itemize}
\item[$(i)$] The two colliding hadrons are represented by valence
quarks carrying their quantum numbers (charges) plus the accompanying
clouds of gluons (which represent also the sea $q\bar{q}$ pairs and
therefore should be regarded as effective ones).
\item[$(ii)$] In the course of a collision the gluonic clouds interact
strongly depositing in the central region of the reaction fractions
$x$ and $y$ of the initial energy-momenta of the respective projectiles
in the form of a gluonic {\it Central Fireball} (CF).
\item[$(iii)$] The valence quarks (plus those gluons which did not
interact) get excited and form {\it Leading Jets} (LJ's) 
which then populate mainly the fragmentation regions of the
reaction.
\end{itemize}
The fraction of energy stored in the CF is therefore equal to
$K\, =\, \sqrt{x y}$ and its rapidity is $Y=\frac{1}{2}\, ln
\frac{x}{y}$. These two quantities provide then a sort of dynamically
calculated initial conditions for any statistical model of
multiparticle production and that was one of the initial aims of the
IGM. 

According to the IGM the CF consists of {\it minifireballs} (MF) formed from
pairs of colliding gluons. In collisions at higher scales a MF
is the same as a pair of minijets or jets. In the study of energy
flow the details of fragmentation and hadron production are not
important. Most of the MF's will be in the central region and we
assume that they coalesce forming the CF. The collisions leading to MF's
occur at different energy scales given by 
$Q^2_i = x_i\,y_i\,s$, where the index $i$ labels a particular kinematic 
configuration where the gluon from the projectile has momentum $x_i$ and 
the gluon from the target has $y_i$. We have to choose the scale where we
start to use perturbative QCD. Many studies in the literature converge to 
the value $Q^2_{min} = p_{T\, min}^2 =  (2.3\,GeV)^2$. Below this
value we have to assume that we can still talk about individual soft
gluons and due to the short correlation length (found in lattice QCD
calculations) between them  they still interact mostly pairwise. In
this region we can no longer use  the distribution functions
extracted from DIS nor the perturbative elementary cross sections. 

The central quantity in the IGM is the probability to form a CF
carrying momentum fractions $x$ and $y$ of two colliding hadrons. It is
defined as the sum over an undefined number $n$ of MF's:  
\begin{eqnarray}
\chi(x,y) & = & \sum_{n_1} \, \sum_{n_2} \cdots \sum_{n_i} \, \delta
  \left[ x - n_1 \, x_1 - \cdots - n_i \, x_i \right] \delta \left[
   y - n_1 \, y_1 - \cdots - n_i \, y_i \right] \cdot P(n_1) \cdots 
    P(n_i) \; \nonumber \\[0.4cm]
& = & \sum_{\{n_i\}} \left\{ \delta \left[ x - \sum_i n_i\, x_i \right]
   \delta \left[ y - \sum_i n_i\, y_i \right] \right\} \prod_{\{n_i\}}
    P(n_i)
\label{eq:isachi} 
\end{eqnarray}

The delta functions in the above expression garantee energy momentum 
conservation and $P(n_i)$ is the probability to have $n_i$ collisions
between gluons with $x_i$ and $y_i$. If the collisions are independent 
$P(n_i)$ is given by:
\begin{eqnarray}
P(n_i) &=& \frac{(\overline n_i)^{n_i} \, exp(-\overline n_i)}{n_i\,!}
\end{eqnarray}

Inserting $P(n_i)$ in (\ref{eq:isachi}) and using the following integral 
representations for the delta functions:
\begin{eqnarray}
&& \delta \left[ x - \sum_i n_i \, x_i \right] \, = \, \frac{1}{2\pi}
     \int^{+\infty}_{-\infty} dt \,\exp \left[ it \left( x - \sum_i
      n_i \, x_i \right)\right] \\[0.4cm]
&& \delta \left[ y - \sum_i n_i \, y_i \right] \, = \, \frac{1}{2\pi}
    \int^{+\infty}_{-\infty} du\, \exp \left[ iu \left( y - \sum_i
     n_i \, y_i \right)\right] 
\end{eqnarray} 
we can perform all summations and products arriving at:
\begin{eqnarray}
\chi(x,y) \, = \, \frac{1}{(2\pi)^2} \int^{+\infty}_{-\infty} dt 
    \int^{+\infty}_{-\infty} du \,\exp [i (tx + uy)] \exp \left\{
     \sum_i \left\{ \overline{n}_i \left[ e^{-i(tx_i+uy_i)} - 1
      \right] \right\} \right\}
\label{eq:yf}
\end{eqnarray}

Taking now the continuum limit:
\begin{eqnarray}
\overline{n}_i \, = \, \frac{d \overline{n}_i}{dx' \, dy'} \,   
    \Delta x' \Delta y'  \;\; \longrightarrow \; \; d\overline{n}
     \, = \, \frac{d\overline{n}}{dx' \, dy'} \, dx' \, dy' 
\end{eqnarray}
we obtain:
\begin{eqnarray}
\chi(x,y) & = & \frac{1}{(2\pi)^2} \int^{+\infty}_{-\infty} dt 
    \int^{+\infty}_{-\infty} du \,\exp[i(tx + uy)] \times \nonumber \\[0.4cm]
   && \times \exp\left\{ \int^1_0 dx' \int^1_0 dy' \, \omega(x',y')
       \left[ e^{-i (tx' + uy')} -1 \right]\right\}
\label{eq:yfs}
\end{eqnarray}
where
\begin{eqnarray}
\omega(x',y')\,=\, \frac{d \overline {n}}{ d x' \, d y'} .
\end{eqnarray}

This function $\omega(x',y')$, sometimes called the spectral function,    
represents the average number of gluon-gluon collisions as a function of 
$x'$ e $y'$, contains all the dynamical inputs of the model and has the 
form:
\begin{equation}
\omega(x',y')\, =\, \frac{\sigma_{gg}(x'y's)}{\sigma(s)}
   \, G(x')\, G(y')\, \Theta\left(x'y' - K^2_{min}\right),
   \label{eq:omega}
\end{equation}
where $G$'s denote the effective number of gluons from the
corresponding projectiles (approximated by the respective gluonic 
structure functions) and $\sigma_{gg}$ and $\sigma$ are the gluon-gluon and
hadron-hadron cross sections, respectively. In the above expression $x'$ and
$y'$ are the fractional momenta of two gluons coming from the projectile and
from the target whereas $K_{min}=m_0 / \sqrt{s}$, with $m_0$  being the
mass of lightest produced state and $\sqrt{s}$ the total c.m.s. energy. 
$m_0$ is a parameter of the model. 

The integral in the second line of eq. (\ref{eq:yfs}) is dominated by the 
low $x'$ and $y'$ region. Considering the singular behaviour of the  $G(x)$ 
distributions at the origin we make the following approximation: 
\begin{eqnarray}
\left[ e^{-i (tx' + uy')} -1 \right]\,\simeq\, -i  (tx' + uy') -\frac{1}{2}
(tx' + uy')^2
\end{eqnarray}

With this approximation it is possible to perform the integrations in
(\ref{eq:yfs}) and obtain the final expression for $\chi(x,y)$ discussed 
in the main text:
\begin{eqnarray}
&&\chi(x,y)\, =\, \frac{\chi_0}{2\pi\sqrt{D_{xy}}}\cdot \nonumber \\
&&\cdot \exp \left\{ - \frac{1}{2D_{xy}}\,\left[
  \langle y^2\rangle (x - \langle x\rangle )^2 +
  \langle x^2\rangle (y - \langle y\rangle )^2 -
  2\langle xy\rangle (x - \langle x\rangle )(y - \langle y\rangle )
  \right] \right\} 
\label{eq:chifin}
\end{eqnarray}
where
\begin{eqnarray}
D_{xy} &=& \langle x^2\rangle \langle y^2\rangle - 
           \langle xy\rangle ^2  \nonumber 
\end{eqnarray}
and
\begin{eqnarray}
\langle x^ny^m\rangle &=& \int_0^1\! dx\,x^n\, \int_0^1\! dy\, 
y^m\, \omega (x,y), \label{eq:defmoms}
\end{eqnarray}
$\chi_0$ is a normalization factor defined by the condition:
\begin{eqnarray}
 \int_0^1\!dx\, \int_0^1\! dy\, \chi(x,y) \theta(xy - K_{min}^2) = 1
\end{eqnarray}

In order to evaluate the distribution  (\ref{eq:chifin}) we need to choose 
the value of  $m_0$,  the semihard scale $p_{T\,min}$ and define  
$G(x)$ and $\sigma_{gg}$ in both interaction regimes.  
We take  $p_{T\,min}\,=\,2.3\,GeV$ and $m_0\,=\, 0.35\,GeV$. These are
the two scales present in the model. The semihard gluon-gluon cross section 
is taken, at order $\alpha_s^2$, to be: 
\begin{eqnarray}
\hat{\sigma}^h_{gg}(x\,,\,y\,,\,s)\,=\, 
\kappa \frac{\pi}{16\,p^2_{T\,min}} \left[
\alpha_s(Q^2)\right]^2\,H
\end{eqnarray}
where
\begin{eqnarray}
H\,=\,36\,T\,+\,\frac{51\,\lambda\, T}{4\,x\,y}\,-\, 
\frac{3\lambda^2\,T}{8\,x^2\,y^2}\,+\,\frac{9\,\lambda}{x\,y}\,
\ln\left[\,\frac{1-T}{1+T}\,\right]
\end{eqnarray}
and
\begin{eqnarray}
T\,=\,\left[1-\frac{\lambda}{x\,y}\right]^{\frac{1}{2}} \,;\, 
\lambda\,=\, \frac{4\,p^2_{T\,min}}{s}
\end{eqnarray}
The parameter $\kappa$ is the one frequently used to incorporate higher 
corrections in $\alpha_s$ and is 
$1.1 \leq \kappa \leq 2.5$ according to the choice of $G(x)$, of the 
scale $Q^2$ and  $p_{T\,min}$. For $p_{T\,min}=2.3\,GeV$, 
$\kappa=2.5$.

The coupling constant is given by:
\begin{eqnarray}
\alpha_s(Q^2)\,=\, \frac{12\,\pi}{\left(33-2 N_f\right)\,
\ln\left[\frac{Q^2}{\Lambda^2}\right]}
\end{eqnarray}
where $\Lambda\,=\,0.2\,GeV$ and $N_f\,=\,3$ is the number of active 
flavors. As usual in minijet physics we choose
$Q^2\,=\,p^2_{T\,min}$ and use the distributions $G(x,Q^2)$
parametrized in literature.

When the invariant energy of the gluon pair $\hat{s}$ is the interval
$m^2_0\,\leq\, \hat{s}\,=\,xys\,\leq\,4p^2_{T\,min}$ we are outside the 
perturbative domain. Parton-parton cross sections in the non-perturbative regime 
have been parametrized in \cite{valin} leading to a successfull quark-gluon model for 
elastic and diffractive scattering.   Recently these non-perturbative cross  
sections have been calculated in the stochastic vacuum model \cite{menon}. The 
obtained cross sections are functions of the gluon condensate and of the gluon 
field correlation  length, both quantities extracted from lattice QCD calculations. 
In order to keep our treatment simple we shall adopt the older parametrization for 
the gluon-gluon cross section used in \cite{valin}: 
\begin{eqnarray}
\hat{\sigma}^s_{gg}(x\,,\,y\,,\,s)\,=\, \frac{\alpha_0} {x\,y\,s} 
\end{eqnarray}
where $\alpha_{0}$ is the second parameter of the model \cite{IGM}.\\

\vspace{1cm}

\underline{Acknowledgements}: This work has been supported by FAPESP, 
under contract 95/4635-0, CNPQ (Brazil) and KBN (Poland).  
F.S.N. would also like to thank the
The Andrzej Soltan Institute for Nuclear Studies, Warsaw, for the
warm hospitality extended to him during his visit there. 
We would like to thank Drs 
E.C. de Oliveira and Z. W\l odarczyk for heplful discussions
on diffractive and cosmic ray experiments.

%\newpage

\vspace{1cm}
%\newpage
\noindent
{\bf Figure Captions}\\
\begin{itemize}
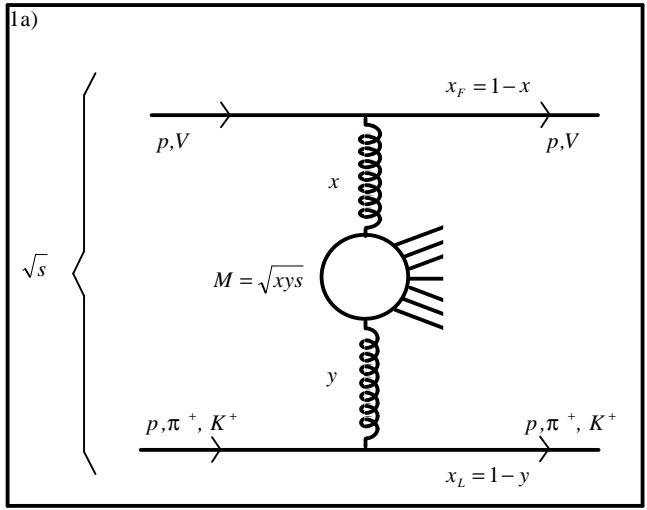
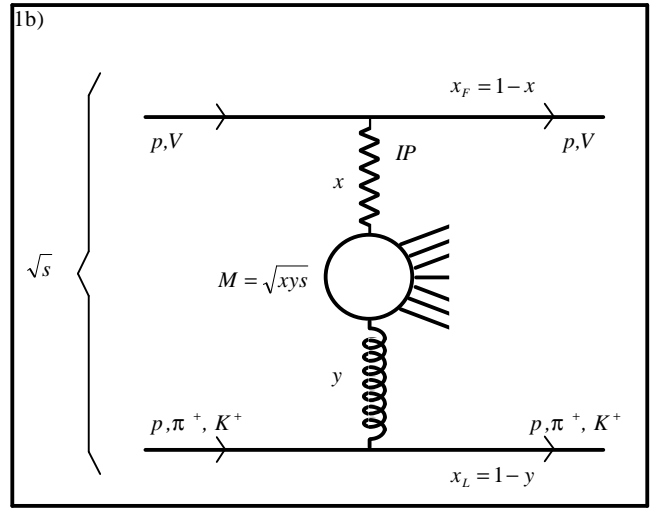
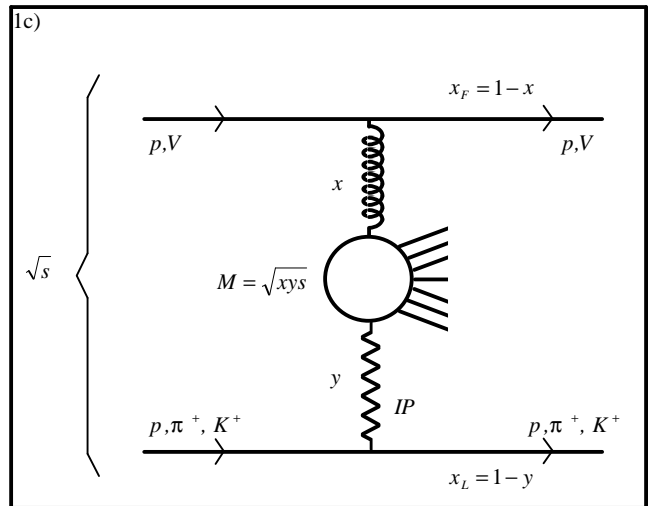
\item[{\bf Fig. 1}] The schematic description of IGM. In Fig. 1a) we
                    show a non-diffractive
                    event.  The upper (lower) leg represents 
                    the leading particle with
                    the momentum fraction $x_F=1-x$ ($x_L=1-y$). 
		    In Fig. 1b)
                    the hadron in the upper leg sends 
                    a $I\!\!P$ with momentum
                    fraction $x$ and remains a leading particle with 
                    momentum fraction $x_F=1-x$. Fig. 1c)
                    corresponds to the case where $I\!\!P$ is emitted
                    from hadron in the lower leg, which keeps momentum 
		    $x_L=1-y$.

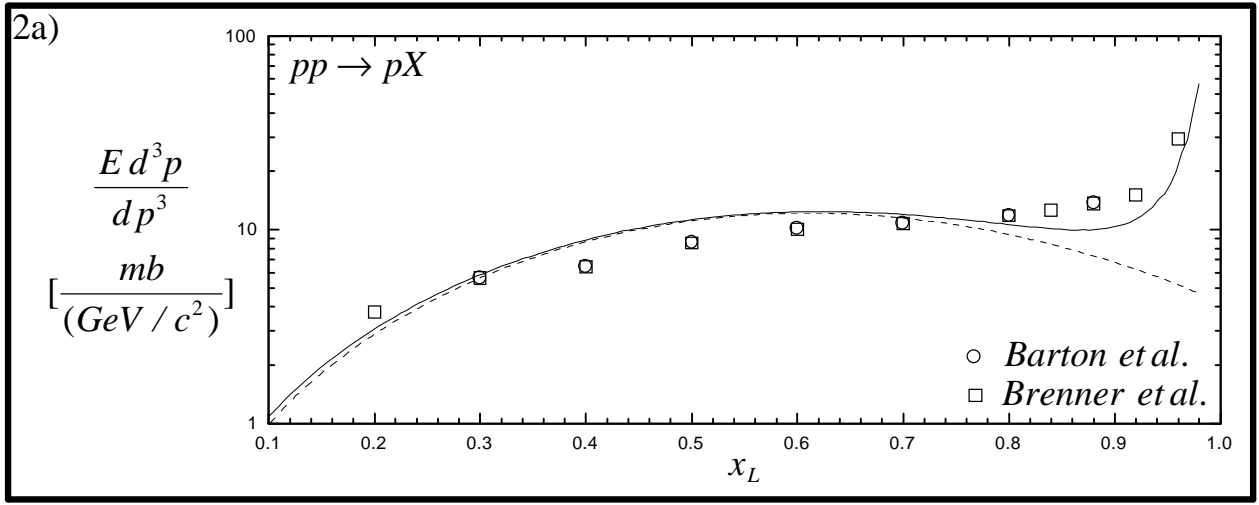
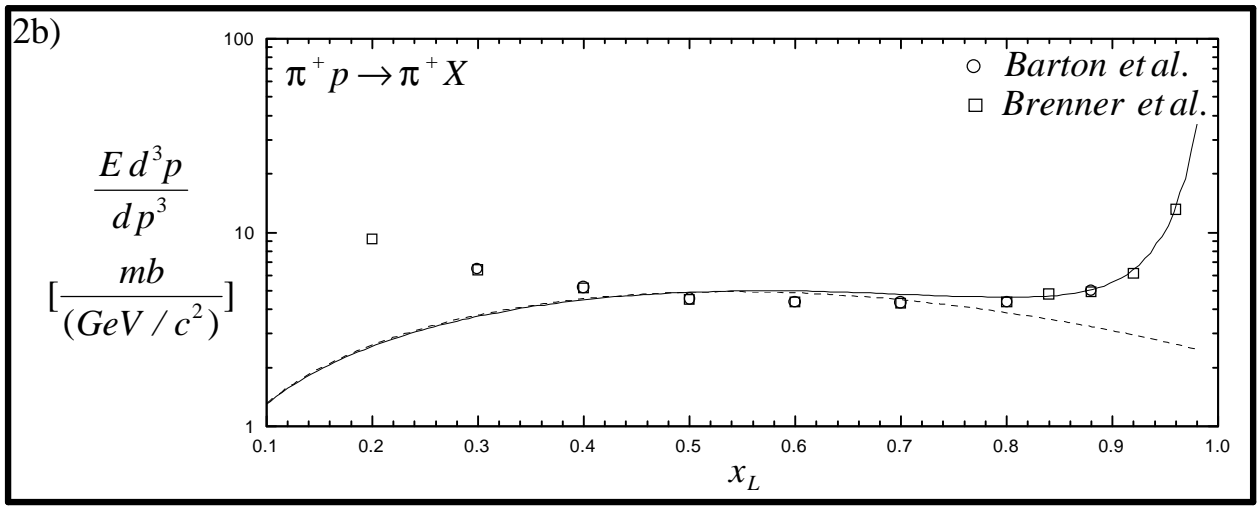
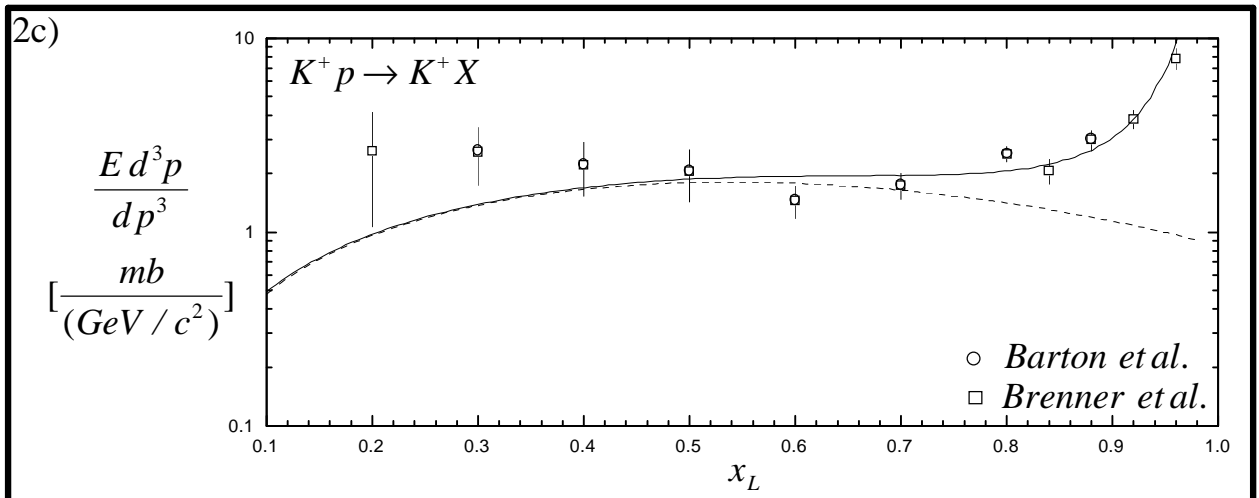
\item[{\bf Fig. 2}] Comparison of our spectra $F(x_L)$ with data from 
                    ref. \cite{DATA} for a) leading protons, b) leading pions 
                    and c) leading kaons. Dashed and solid lines show  the 
                    non-difractive component and the total curve respectively.
                                
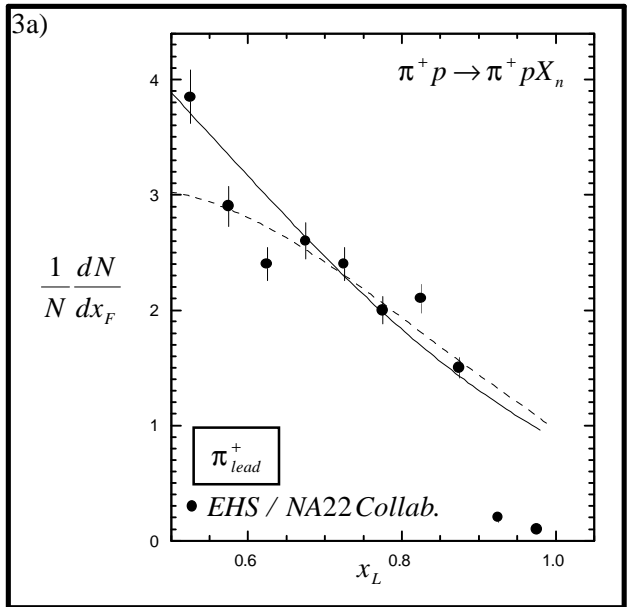
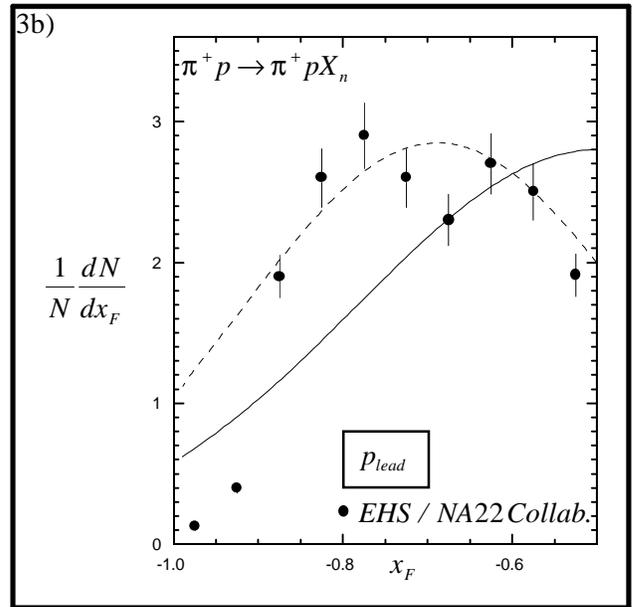
\item[{\bf Fig. 3}] a) Comparison of our spectra $F(x_L)$ for leading pions 
                    with data from ref. \cite{EHS} in the reaction 
                    $\pi^{+} + p \rightarrow \pi^{+} + p + X$. Solid and dashed
                    lines correspond to the choices $m_0=0.35\,GeV$ and 
                    $m_0=0.45\,GeV$ respectively. b) the same as a) for the 
                    leading proton spectrum $F(x_F)$ measured in the same
                    reaction.
                    
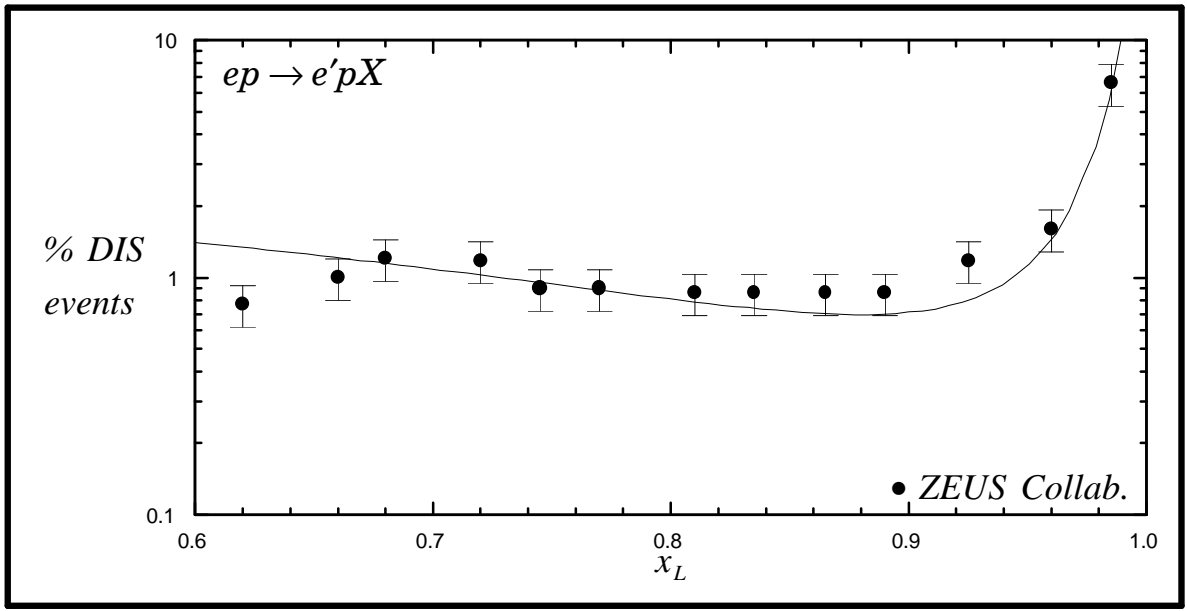
\item[{\bf Fig. 4}] Comparison between our calculation and the LP spectrum
                    measured at HERA by the ZEUS collaboration, 
                    ref. \cite{Cart}.

\end{itemize}


\begin{thebibliography}{99}

\bibitem{DATA} D.S. Barton {\sl et al.}, {\sl Phys. Rev.} {\bf D27} 
               (1983) 2580; A.E. Brenner {\sl et al.},  
               {\sl Phys. Rev.} {\bf D26} (1982) 1497. 


\bibitem{EHS} EHS/NA22 Collaboration, N.M. Agababyan {\sl et al.}, 
	      {\sl Z. Phys.} {\bf C75} (1996) 229.


\bibitem{Cart} N. Cartiglia, {\it Leading Baryons at Low $x_{L}$ in
               DIS and Photoproduction at ZEUS}, hep-ph/9706416.


\bibitem{VDM} S.D. Holmes, W. Lee and J.E. Wiss, {\sl Ann. Rev. Nucl. Part.
              Sci.} {\bf 35} (1985) 397; T.H. Bauer, R.D. Spital,
              D.R. Yennie and F.M. Pipkin, {\sl Rev. Mod. Phys.} {\bf
              50} (1978) 261.


\bibitem{FGS} G.M. Frichter, T.K. Gaisser and T. Stanev, 
              {\sl Phys. Rev.} {\bf D56} (1997) 3135.


\bibitem{CR} Cf., for example, A.A. Watson, {\sl Nucl. Phys.} 
             {\bf B} {\sl (Proc. Suppl.)} {\bf B60} (1998) 171;
             J.W. Cronin, {\sl Nucl. Phys.} {\bf B} {\sl (Proc. Suppl.)}
             {\bf B28} (1992) 213; {\it The Pierre Auger Project},
             Design Report, November 1996 (Second edition),
             http://www-td-auger.fnal.gov:82

\bibitem{QGP} Cf., for example, {\it Quark Matter' 96}
              proceedings in: {\sl Nucl. Phys.} {\bf A610} (1996)
              and references therein.


\bibitem{BER} A. Berera {\sl et al.}, {\sl Phys. Lett.} {\bf B403}
             (1997) 1.


\bibitem{WAL} W.D. Walker, 
              {\sl Phys. Rev.} {\bf D53} (1996) 1886.


\bibitem{charm} F.O. Dur\~aes, F.S. Navarra, C.A.A. Nunes and 
                G. Wilk, {\sl Phys. Rev.} {\bf D53} (1996) 6136.

\bibitem{IGM} G.N. Fowler, F.S. Navarra, M. Pl\"umer, 
              A. Vourdas, R.M. Weiner and G. Wilk, {\sl Phys.Rev.} 
              {\bf C40} (1989) 1219. 

\bibitem{DNW} F.O. Dur\~aes, F.S. Navarra and G. Wilk, {\sl Phys.
              Rev.} {\bf D47} (1993) 3049 and {\bf D50} (1994) 
              6804.

\bibitem{IGM97} F.O. Dur\~aes, F.S. Navarra and G. Wilk, {\sl Phys.
                Rev.} {\bf D55} (1997) 2708.

\bibitem{HERA} F.O. Dur\~aes, F.S. Navarra and G. Wilk, {\sl Phys.
               Rev.} {\bf D56} (1997) R2449.


\bibitem{PSI} F.O. Dur\~aes, F.S. Navarra and G. Wilk, 
              {\it Leading Particle Effect in $J/\Psi$ Elasticity Distribution}, 
              hep-ph/9803325 and {\sl Mod. Phys. Lett.}
              {\bf A} (1998), in press.

\bibitem{wang}  X. N. Wang and M. Gyulassy, {\sl Phys. Rev.} 
                {\bf D44} (1991) 3501; {\bf D45} (1992) 844.


\bibitem{gia} A. di Giacomo and H. Panagopoulos, {\sl Phys. Lett.} 
              {\bf B285} (1992) 133.


\bibitem{brown} N. Brown, {\sl Mod. Phys. Lett.} {\bf A4} (1989) 2447. 


\bibitem{gs}  T.K. Gaisser and T. Stanev, {\sl Phys. Lett.} {\bf B219} 
              (1989) 375.

\bibitem{sj}  T. Sjostrand and M. van Zijl, 
              {\sl Phys. Rev.} {\bf D36} (1987) 2019.


\bibitem{lupia} S. Lupia, W. Ochs and  J. Wosiek, 
                {\it Poissonian limit of soft gluon multiplicity}, 
                hep-ph/9804419.


\bibitem{HAMA} S. Paiva, Y. Hama and T. Kodama, 
               {\sl Phys. Rev.} {\bf C55} (1997) 1455; S. Paiva and 
               Y. Hama, {\sl Phys. Rev. Lett.} {\bf 78} (1997) 3070.


\bibitem{BRAN} A. Brandt {\sl et al.}, {\it A forward proton detector at D0},
               FERMILAB-PUB-97/377 (1997) and E. Oliveira, 
               private communication.
               

\bibitem{zeusp} ZEUS Collaboration, M. Derrick {\sl et al.}, hep-ex/9804013.

\bibitem{cov} R.J.M. Covolan and  M.S. Soares,  
              {\sl Phys. Rev.} {\bf D57} (1998) 180.

\bibitem{zeus} ZEUS Collaboration, M. Derrick {\sl et al.}, {\sl Phys. Lett.} 
               {\bf B356} (1995) 129; 
               {\sl Z. Phys.} {\bf C68} (1995) 569 and 
               references therein.

\bibitem{h1} H1 Collaboration, T. Ahmed {\sl et al.},  {\sl Nucl. Phys.} 
             {\bf B435} (1995) 3; {\bf B429} (1994) 477.


\bibitem{CRINEL} For example, in the cosmic ray experiments it is usually
                 assumed that $K_{\pi N} = 1.5\, K_{pp}$, which
                 is traced to analysis of data like those by
                 G. Donaldson {\sl et al.}, {\sl Phys. Lett.} {\bf B73}
                 (1978) 375, performed in terms of the additive quark
                 models (cf., E. Fishbach and G.W. Look, {\sl Phys. Rev.}
                 {\bf D15} (1977) 2576) - private information from 
                 Z. W\l odarczyk.

\bibitem{SNS} A. Szczurek, N.N. Nikolaev and J. Speth, 
              {\it Leading proton spectrum from DIS at HERA}, 
              hep-ph/9712261.


\bibitem{AH}  ZEUS Collaboration, M. Derrick {\sl et al.}, 
              {\sl Phys. Lett.} {\bf B384} (1996) 388.



\bibitem{FOOT} One should mention here that there is another possible
               difference between nucleons and mesons which can contribute
               to the different behaviour of the leading particles in
               both cases. It is connected with the triple gluon
               junction present (in some models) in baryons but not
               in mesons, which, if treated as elementary object, can
               influence sunbstantially LP spectra (cf. D. Kharzeev,
               {\sl Phys. Lett.} {\bf B378} (1996) 238 and references 
               therein). We shall not discuss this possibility in this paper.


\bibitem{geiger} K.Geiger, {\sl Phys. Rev.} {\bf D46} (1992) 4965 and
                 {\bf 46} (1992) 4986.

\bibitem{grv} M.Gl\"uck, E.Reya and A.Vogt, {\sl Z. Phys.} {\bf C47}
              (1995) 433; A.Edin and G.Ingelman, {\it A model for the
              parton distributions in hadrons}, TSL/ISV-98-0194
              and DESY 98-035 ( hep-ph/9803496 ).

\bibitem{valin} P. L'Hereux, B. Margolis and P. Valin, {\sl Phys. Rev.} 
                {\bf  D32} (1985) 1.

\bibitem{menon} A.F. Martini, M.J. Menon and D.S. Thober, {\sl Phys. Rev.} 
                {\bf  D57} (1998) 3026.


\end{thebibliography}
\end{document}